\documentclass{icrc2009}

\usepackage{graphicx}   % for including figures
\usepackage[caption=false]{caption}    % for captions
\usepackage[font=footnotesize]{subfig} % subfig.sty for a double column floating figure using two subfigures
\usepackage{fixltx2e}
\usepackage{url}

\newcommand{\shorttitle}[1]%
{\markboth{Proceedings of the 31\MakeLowercase{$^{st}$} ICRC, {\L}\'{o}d\'{z} 2009}{#1} }
\newcommand{\etal}{\MakeLowercase{\textit{et al. }}} % "et al."

\begin{document}
\title{MARS, the MAGIC Analysis and Reconstruction Software}

\author{\IEEEauthorblockN{
    Abelardo Moralejo\IEEEauthorrefmark{1},
    Markus Gaug\IEEEauthorrefmark{2}, 
    Emiliano Carmona\IEEEauthorrefmark{3},
    Pierre Colin\IEEEauthorrefmark{3},
    Carlos Delgado\IEEEauthorrefmark{2}, \\
    Saverio Lombardi\IEEEauthorrefmark{4},
    Daniel Mazin\IEEEauthorrefmark{1},
    Villi Scalzotto\IEEEauthorrefmark{4},
    Julian Sitarek\IEEEauthorrefmark{3}\IEEEauthorrefmark{5},
    Diego Tescaro\IEEEauthorrefmark{1} \\
    for the MAGIC collaboration}
  \\
\IEEEauthorblockA{\IEEEauthorrefmark{1}IFAE, Edifici Cn, Campus UAB,
  E-08193 Bellaterra, Spain}
\IEEEauthorblockA{\IEEEauthorrefmark{2}Instituto de Astrof\'{\i}sica de Canarias,
  E-38200 La Laguna, Tenerife, Spain }
\IEEEauthorblockA{\IEEEauthorrefmark{3}Max-Planck Institut f\"ur
  Physik, D-80805 M\"unchen, Germany }
\IEEEauthorblockA{\IEEEauthorrefmark{4}Universit\`a di Padova and INFN,
I-35131 Padova, Italy}
\IEEEauthorblockA{\IEEEauthorrefmark{5}{University of \L\'od\'z,
    PL-90236 \L\'od\'z, Poland}}
}

% please write the preseter's name and short title (3-4 words maximum)
%    which will appear at the header of the even pages.
\shorttitle{Abelardo Moralejo \etal MARS, the MAGIC Analysis Software}
\maketitle

\begin{abstract}
With the commissioning of the second MAGIC gamma-ray Cherenkov
telescope situated close to MAGIC-I, the standard analysis package of
the MAGIC collaboration, MARS, has been upgraded in order to perform the
stereoscopic reconstruction of the detected atmospheric showers. MARS
is a ROOT-based code written in C++, which includes all the necessary
algorithms to transform the raw data recorded by the telescopes into
information about the physics parameters of the observed targets. An
overview of the methods for extracting the basic shower parameters is
presented, together with a description of the tools used in the
background discrimination and in the estimation of the gamma-ray
source spectra.
\end{abstract}

 
\section{Introduction}
MARS ({\it MAGIC Analysis and Reconstruction Software}) is a collection
of ROOT-based \cite{root} programs written in C++ 
for the analysis of data from gamma-ray Cherenkov telescopes. MARS has
been developed during the last decade within the MAGIC collaboration,
and is currently the official analysis package of MAGIC, an Imaging
Atmospheric Cherenkov Telescope (IACT) located on the island of La
Palma (Spain). A second MAGIC telescope, which is presently in its
commissioning phase, will soon allow to carry out observations in
stereoscopic mode, thus enhancing significantly the performance of the
instrument \cite{stereo}. MARS is also being used in the analysis of
Monte Carlo - simulated data of large arrays of IACTs, aimed at the
study of the possible configurations of the next-generation
ground-based gamma-ray observatory dubbed CTA (Cherenkov Telescope
Array \cite{cta}).
\par
The data analysis chain implemented in MARS is divided into several
steps, each of which is performed by an independent program which
takes as input the output of one or more of the previous
stages. The initial input to MARS are the raw data recorded by the
telescopes, consisting of binary files containing the full information
available per pixel (digitized signal amplitude vs. time) for every
triggered event, plus ascii files containing regular reports 
from the different telescope subsystems (like the telescope drive, the
trigger system or the weather station). Throughout the analysis chain,
the data are organized in ROOT trees containing a set of ``parameter
containers'' for every entry. Typically, the core of a MARS program
\cite{tsukuba03} is an event loop which executes an ordered list of
tasks (the ``task list'') for every event in the input file. Besides
the task list, every loop is associated to a ``parameter list''. It
contains pointers to all the parameter containers holding the input
data needed by the tasks, and to the containers where the tasks
store the results of their calculations.

\section{Signal extraction and calibration}
The first program in the analysis chain is called {\it callisto}, and
its main purpose is to calibrate the raw data. After subtraction of
the pedestal offsets, several algorithms are available in MARS for
extracting the signal of each pixel. Since the upgrade of the data
acquisition to 2 Gsample/s in early 2007, the 
standard method has become the integration around the peak of a cubic
spline built from the raw digitized pulse. Besides the integrated
signal, the arrival time of the pulse is computed, as the position of
the rising edge of the spline at 50$\%$ of the peak value. {\it
  Callisto} then equalizes the response of the different camera pixels
to account for differences in gain 
(flatfielding), and introduces relative offsets to correct for
deviations in signal arrival times. For these purposes, {\it callisto}
makes use of dedicated pedestal and calibration runs, and also of
pedestal and calibration events interleaved with the ordinary data,
which help to track possible drifts in the pedestal baseline and in the
gains. An absolute calibration procedure, based on the F-factor
method \cite{ffactor}, is also applied to convert the reconstructed
signal amplitudes into physically meaningful units (photoelectrons).

\section{Image cleaning and parametrization}

After calibration, the next step in the analysis chain is the
parametrization of each shower image by a small set of parameters
which describe in a compressed way its orientation, shape and timing
properties. Among these quantities are the Hillas parameters, which
are basically the moments up to second order of the light distribution
on the camera. Obviously, before calculating the moments of the light
distribution, a ``cleaning'' has to be performed in order to remove
pixels which most likely do not contain light from the shower, and whose
``signals'' are just the result of the fluctuations of the light of
the night sky. These tasks are performed in MARS by the program called
{\it star}. 
\par
In {\it star}, the arrival time of the light at each pixel is used
along with the signal amplitude both in the cleaning 
procedure and in the calculation of some of the image
parameters \cite{timing}. The decision to accept a certain pixel as
part of the image relies on the strength of the signal and on its
contemporaneity with those in neighboring pixels. As for time-related
image parameters, it has been shown recently \cite{timing} that the
evolution of the arrival time of the light along the major axis of the
shower image can be used to improve the performance of MAGIC-I
operated as a standalone Cherenkov telescope, allowing to halve the
rate of residual background events.
\par
During this stage of analysis, the ring-shaped images from isolated
muons are identified, and their brightness and broadness are analyzed
in order to provide information on the overall light
collection efficiency of the telescope and on the optical point spread
function of the mirror dish. This information is needed to tune the
Monte Carlo simulation used for later analysis. 

\section{Stereoscopic shower reconstruction}

Up to the image parametrization ({\it callisto} and {\it star}), the
MARS analysis chain runs over the data of each telescope
separately. At this point, we have two sets of {\it star} files, one
per telescope, which contain two different views of the same
showers\footnote{As of now, the files also contain showers which
are not seen by the other telescope, but this will not be the case
once the inter-telescope coincidence trigger is implemented.}. A
program called {\it superstar} reads in the two streams of
files and identifies the matching pairs of events. It then
calculates the parameters 
which define the shower axis (direction and impact point) from the simple
intersection of two planes, each of them defined by one of the
recorded images (plus the position and orientation of the
corresponding telescope). In the case of a 2-telescope system like
MAGIC, there is only one solution for the geometry of the shower axis,
and its accuracy depends on the relative positions of the telescopes
and the shower: the more parallel the two images on the camera planes
are, the larger the uncertainties in the reconstructed parameters. As
of now, only images with a relative angle of at least $30^\circ$ are
used in the stereo analysis, but work is going on to improve the
reconstruction of the rest of events through the analysis of image
shapes (along the lines of the {\it DISP} method \cite{disp}), which
constrain the distance between the image center of gravity and the
point on the camera which corresponds to the shower direction.
\par
Once the shower axis is determined, an estimate of the height of the
shower maximum is made from the angle at which the image center of
gravity is viewed from each telescope. {\it Superstar} also calculates
the impact parameter of the shower with respect to each telescope, and
obtains an estimate of the energy of the primary (assuming
it is a gamma-ray) by using simple Monte Carlo - generated lookup
tables of the energy versus image Size, impact parameter, atmospheric
depth of the shower maximum and zenith angle.

\section{Background discrimination}

The standard procedure in MARS to suppress the unwanted background
showers produced by charged cosmic rays makes use of a multivariate
classification method known as Random Forest (RF)
\cite{randomforest}. For every event, the algorithm takes as input a
set of image parameters, and produces one single parameter as output,
called {\it hadronness}, which is in 
the range from 0 to 1. A low value of {\it hadronness} indicates the event is
a good gamma candidate. Only events with {\it hadronness} below a certain
cut value will be used for the subsequent steps of the analysis. The
MARS program in charge of the learning phase of the RF is called {\it
  osteria}, which takes as input a set of {\it star} files from Monte
Carlo gamma rays and another one of real MAGIC data from observations
of a sky region devoid of any gamma-ray source (hence containing
almost exclusively background events). The RF can take as input
parameters both global 
shower parameters from the stereo reconstruction (like the 
height of the shower maximum or the estimated energy) and 
image parameters of each one of the telescopes (e. g. the Hillas
parameters).
\par
It must be noted that the RF method can be used not only to
classify events into different populations, but also to estimate the
value of an unkown continuous quantity, like the energy of the primary
gamma-ray, which is correlated with the RF input parameters. This is
in fact the default method for energy estimation that we have been
using in the MARS analysis of single-telescope (MAGIC-I) observations.

\section{Light curve and energy spectrum}

The differential energy spectrum of the observed gamma-rays is
estimated by the {\it fluxlc} program of the MARS package. After a cut
in {\it hadronness} ($< h_{max}$), the gamma-ray excess is determined
by counting all events within an angular distance $\theta_{max}$ of
the source direction\footnote{In single-telescope observations,
the simple orientation of the shower image with respect to the nominal
source position on the camera (ALPHA parameter) is normally used
instead of $\theta$}, and subtracting from it an estimate of the
number of background events. For the case of {\it wobble} observations
(in which the telescope is pointed $0.4^\circ$ away from the source), the
so-called false-source method \cite{disp}, can be used for background
estimation. For ON-source observations (for which the candidate source
is located in the center of the camera), an additional sample of OFF
data (with no gamma-ray source on the field of view) is needed for
this purpose. 
\par
The excess of events is obtained in bins of estimated
energy and divided by the total effective observation time
$t_{\mathrm{eff}}$ and by the gamma-ray effective area for each energy bin
$A_{\mathrm{eff}}$. The effective area after all cuts is calculated
using a test sample of Monte Carlo gamma-rays (which must be
statistically independent of the Monte Carlo sample used by {\it
  osteria} for training the background suppression). 
\par
The effective
area depends on the direction in local coordinates of the observed
gamma-rays, mainly due to the variation with the zenith distance ({\it
  Zd}) of the air mass along the pointing direction, but also due to
magnetic field effects and to the relative orientation of the shower
and the system of two telescopes, which introduce a dependence on the
azimuth ({\it Az}) as well. The effective area in an energy bin is
therefore obtained as a weighted average $\Sigma A_{\mathrm{eff}}(Az,
Zd) \times w(Az, Zd)$, where the weights $w$ are proportional to the
observation times spent in each bin of azimuth and zenith traversed by
the source. 
\par
Different cuts $h_{max}$ and $\theta_{max}$ are tried in the standard
analysis to test the stability of the derived spectrum: a significant
variation of the result would indicate that the cut efficiencies (and
hence the effective areas) are not well reproduced by the Monte Carlo
simulation, therefore casting doubts on the reliability of the
measurement\footnote{If no such large variation is observed, the small
  changes of the spectrum for different cut efficiencies are
  incorporated into the systematic uncertainty.}.
\par
The errors on the spectral points include statistical errors on the
number of 
excess events and the uncertainty of the effective area. 
The spectrum at this point is calculated in bins of estimated energy
and might, therefore, differ from the true gamma-ray energy
spectrum especially around the energy threshold. The calculation of
the energy spectrum in bins of true energy, a procedure which is called 
unfolding, is presented in the following.
\par
Measurements of the gamma-ray energy are systematically distorted
due to the fact that the detectors are not ideal (e.g.\ have finite
resolution) and the true energy is not directly measured. The
distortions due to biases and finite resolution can be written in the
form: 
\begin{eqnarray}
  && Y(y) = \int M(y,x)S(x)\mathrm{d}x \hspace{0.5cm} \mathrm{or}\nonumber \\
  && Y_i = \sum_j M_{ij}S_j  \hspace{0.5cm}
  \mathrm{or} \hspace{0.5cm} Y = M \times S
\end{eqnarray}
where $y$ is the estimated energy, $x$ is the true energy, $M$
describes the detector response 
(the migration matrix), $Y$ is the measured distribution and $S$ is
the true undistorted distribution. The aim is to determine $S$, given
$Y$ and $M$. There are various approaches to solve this problem.

One of the solutions (called deconvolution) is to invert the matrix
$M$. Although technically correct, this is often useless due to large
correlations between adjacent bins, which imply large 
fluctuations of their contents. This fact is the basis of the
unfolding methods with regularization \cite{regularization}.
In these methods one considers two terms: one term, $\chi_{0}^{2}$, expressing
the degree of agreement between the prediction $M \times S$ and the measurement $Y$, and another term,
$Reg$, which is a measure of the smoothness of $S$. A solution for $S$ is obtained by minimizing the expression
\begin{equation}
  \chi^2 = \frac{w}{2} \times \chi^2_0  +  Reg
\end{equation}
for a fixed regularization parameter $w$. Large values of $w$, corresponding to no
regularization, often produce noisy unfolded distributions that perfectly fit the data.
Very small values of $w$ will, on the other hand, overemphasize the regularization, leading
to larger deviation from the measurement but a very smooth unfolded distribution. So, the proper
choice of $w$ is very important. In the MAGIC software, a variety of methods is available
which differ in the way regularization is implemented (see for instance
\cite{tikhonov, bertero}). 

Another approach consists in assuming a parametrization of the true distribution $S$ and then
comparing $M \times S$ with the measured distribution $Y$.
This is called \textit{Forward unfolding}. The main difference to the previous methods is that
an assumption about the true distribution has to be made. Moreover, no explicit regularization
is done in the \textit{Forward unfolding}. On the other hand, the result of the
\textit{Forward unfolding} is just the best fit with corresponding
errors using the {\it a priori} assumed parametrization, but no
spectral points scattered around the unknown real distribution can be
provided. In the MARS analysis, the various unfolding methods are used
for each observation, and the consistency of the results is
checked. Only when results of different unfolding methods agree, the
result of the unfolding is considered trustworthy.  

\par
For the estimation of light curves (gamma-ray flux - in a given energy
range - versus time), no
unfolding of the type just described is used. A simple correction
factor is applied to the effective area in the selected energy
range, to account for the spillover of events with true energies
outside it, under the assumption of a given spectral shape (usually
of power-law type) for the energy spectrum.

\section{Sky map}

In cases in which the exact location of a gamma-ray source is not
known in advance, a blind search in the whole field of view of the
telescope(s) can be performed within MARS using the program {\it celestina}.
\par
The reconstructed directions of all events (in camera coordinates)
surviving the {\it hadronness} cut are converted into celestial
coordinates (Right Ascension, Declination) according to the pointing
position of the telescope and the time stamp of each event. A 
sky map of reconstructed gamma-ray incidence directions is thus
obtained, which contains also the residual background of cosmic ray
events. In order to subtract it, we need to know the camera acceptance
for the background events, and project it on the sky in exactly the
same way, i.e. using the same projection functions, and with the same
telescope orientations and time stamps used for the observation being
analyzed. 
\par
The most general way to obtain the acceptance for the background is to
use an OFF observation (no source in the field of
view). Alternatively, for the search for point-like sources of unknown 
(or uncertain) location within the telescope field of view, one can
use the same data runs containing the signal to estimate the
background acceptance with one of the following methods:

1) If the data are taken in wobble-mode, the camera is divided into
two halves, out of which one half always contains the candidate source
and the second half the false-source (i.e., the position on the
camera opposite to the source with respect to the camera center),
whenever the telescope is pointing at the wobble direction "W1". In the case
of the wobble pointing "W2" (which is defined such that, on the sky,
the assumed source location is the middle point between W1 and W2),
the opposite assignment is used. In this way two signal and two
background halves are obtained which are normalized to each other
using the relative observation time. Obviously for this method we need
to know the {\it approximate} location of the gamma-ray source (within
0.1$^\circ$ or so), since we need to know which camera half contains
only background at any time.

2) A different method starts by filling a 2-D histogram of the
reconstructed arrival directions in camera coordinates. Subsequently,
two new histograms, $h_1$ and $h_2$, are obtained by folding it with a 
Gaussian distribution of width $\sigma \geq \sigma_{PSF}$ (being
$\sigma_{PSF}$ the width of the gamma-ray point-spread function of the
telescope), and $\sigma \times \sqrt{2}$ respectively. A first
estimate of the acceptance for the background is obtained
as~\cite{wavelets} 
\begin{equation} h_r(x,y)=2\times h_2(x,y) - h_1(x,y) \label{eq:conv}
\end{equation}
Assuming a point like gamma-ray source at an unkown
instantaneous\footnote{Given the telescopes have alt-azimuthal mounts,
the sky image on the camera rotates around its center during observations.}
position in the camera, its effect can be approximated by a Gaussian
distribution with a width $\sigma_{PSF}$. Thus the
instantaneous arrival directions at camera position $x,y$ are given by 
\begin{equation}
h(x,y) = S\cdot G(x-x_0,y-y_0|\sigma_{PSF}^2)+B(x,y) \label{eq:conv 1}
\end{equation}
where $S$ is the instantaneous rate of gamma-rays from the 
source, $B(x,y)$ is the instantaneous background rate at camera
position $(x,y)$ and $x_0,y_0$ is the gamma-ray source
position. Additionally the function $G(x,y|\sigma^2)$ is a bi-dimensional
Gaussian distribution  
centered at $(0,0)$ with correlation matrix $I\times\sigma^2$, being $I$ 
the 2x2 identity matrix. In addition we assume that $B(x,y)$ can be described as
the sum of a set of bi-dimensional Gaussians, all with a correlation distance
larger than a given $\sigma_b$ fulfilling $ \sigma^2_b \gg \sigma^2_{PSF}$. Under
those conditions and taking $\sigma=\sigma_{PSF}$, eq. \ref{eq:conv} is approximated by  
\begin{eqnarray}
h_r(x,y) \simeq S \cdot (2\cdot G(x-x_0,y-y_0|3 \sigma^2_{PSF})- \nonumber \\ 
-G(x-x_0,y-y_0|2\sigma^2_{PSF}))+ \nonumber \\
+B(x,y) 
\label{eq:conv 3}
\end{eqnarray}  
where the approximation error is of order $\sigma^2_{PSF}/\sigma^2_b$,
which we experimentally found to be $\sim 0.1$.
The subtraction of Gaussian functions in eq. \ref{eq:conv 3} is
bounded from above by $0.18\cdot G(0,0|\sigma^2_{PSF})$ close to the instantaneous gamma-ray
position and goes to zero exponentially as $G(x-x_0,y-y_0|3 \sigma^2_{PSF})$ far
from the source position. Taking this into account and comparing equation \ref{eq:conv 1}
and \ref{eq:conv 3} we conclude that the latter is a good approximation for the background 
for skymap estimates because of the  signal suppression due to the subtraction
term in eq. \ref{eq:conv 3}.

Finally, to correct the residuals distortions of the background due to
the folding procedure, a factor depending on the distance to the
camera center is applied which guarantees that locally the
normalization is the correct one. Additionally, this correction  
introduces a further suppression factor at the source position. 
\par
For a strong source, this method of estimating the background
acceptance overestimates the background rate (and hence underestimates
the signal rate) at the source position, but is anyway sufficient for
the purpose of estimating the location of the gamma-ray emission and
for blind searches of point-like sources in the FoV.

\section{Summary}
An overview of the methods implemented in the MARS package for the
analysis of data from the MAGIC Cherenkov telescopes has been
presented. The analysis chain is ready to process the stereoscopic
data which will become available once the second MAGIC telescope
becomes fully operational. Results on the expected performance based
on Monte Carlo simulations are presented in a separate contribution
\cite{stereo}.

\end{document}